\documentclass{PoS}

\title{Nucleon Magnetic Moments \\
and \\
Electric Polarizabilities}

\ShortTitle{Nucleon moments and polarizabilities}

\author{W.~Detmold\\
        Department of Physics\\
        College of William and Mary\\
        Williamsburg, VA 23187-8795, USA\\
        {\rm and}\\
        Thomas Jefferson National Accelerator Facility\\
        Newport News, Virginia 23606, USA\\
	E-mail: \email{wdetmold@wm.edu}
}

\author{\speaker{B.~C.~Tiburzi}\\
        Maryland Center for Fundamental Physics\\
        Department of Physics\\
        University of Maryland\\
        College Park, MD 20742-4111, USA\\
        E-mail: \email{bctiburz@umd.edu}
        }

\author{A.~Walker-Loud\\
        Department of Physics\\
        College of William and Mary\\
        Williamsburg, VA 23187-8795, USA\\
        E-mail: \email{walkloud@wm.edu}
        }

\abstract{
Electromagnetic properties of the nucleon are explored with lattice QCD using a novel technique. 
Focusing on background electric fields,
we show how the electric polarizability can be extracted from nucleon correlation functions. 
A crucial step concerns addressing contributions from the magnetic moment, 
which affects the relativistic propagation of nucleons in electric fields. 
By properly handing these contributions, 
we can determine both magnetic moments and electric polarizabilities. 
Lattice results from anisotropic clover lattices are presented. 
Our method is not limited to the neutron;
we show results for the proton as well.
}

\FullConference{The XXVIII International Symposium on Lattice Field Theory, Lattice2010\\
		June 14-19, 2010\\
		Villasimius, Italy}

\begin{document}  

\section{Overview}

Electromagnetic properties of hadrons allow us to glean information about their internal structure. 
Nucleon magnetic moments are well determined experimentally, 
and their deviation from Dirac's value for point-like particles,
$\mu = Q e / 2 M$, 
already implies the composite nature of the nucleon.
The na\"ive quark model provides the prediction:
$\mu_n / \mu_p = - 2/3$;
which, 
while well satisfied experimentally,
results from an uncontrolled approximation.     
Decades after the development of QCD as a fundamental theory of quarks and gluons, 
we are beginning to understand the electromagnetism in the nucleon from first principles. 
A further window to the internal dynamics of the nucleon is given by the study of multipole polarizablities.
These quantities encode the deformation of the nucleon in applied electromagnetic fields, 
and can be accessed experimentally through Compton scattering. 
Near the chiral limit, 
nucleon polarizabilities arise from the deformation of the charged pion cloud, 
and are highly constrained by effective interactions that emerge in the low-energy limit of QCD. 
Lattice QCD will play a crucial role in validating this low-energy picture, 
and the electromagnetic polarizabilities provide an area in which the lattice will influence phenomenology.

We report on our progress in the investigation of hadronic electromagnetic properties using the background field method. 
Previously, 
we have shown how to determine the electric polarizabilities of pseudoscalar mesons using lattice QCD~\cite{Detmold:2009dx}.
Recently, 
we have extended our analysis to the nucleon~\cite{Detmold:2010dx}, 
and this work is our primary concern here.  
We will begin by reviewing the physics of spin-$1/2$ particles in electric fields. 
In particular, 
we treat the spin relativistically in order to account for all effects at second order in the strength of the applied electric field. 
To determine the electric polarizability, 
we show that a background field analogue of the Born subtraction is necessary. 
We devise such a method, 
and apply it to the analysis of nucleon correlation functions calculated using lattice QCD in background electric fields.  
Finally we provide an outlook which describes refinements that must be made in order to confront experiment.

\section{Nucleon in Electric Fields}

Consider a neutron in an external electric field in Minkowski space. 
Because the neutron is composite, 
the external field interacts with the neutron through a tower of non-minimal couplings. 
For fields sufficiently weak compared to the QCD scale, 
$| \vec{E} | / \Lambda^2_{QCD} \ll 1$, 
the tower of terms can be ordered according to a power counting scheme.
Operators with the fewest field-strength tensors and fewest derivatives are the most relevant
in a low-energy effective field theory, 
while those with more field-strength tensors and/or more derivatives are power-law suppressed.%
\footnote{ 
This theory of the neutron and photons is basically the simplest effective field theory one could imagine, 
however, 
there is a subtlety. 
In Minkowski space, 
there are additional non-perturbative effects that are absent in the single-neutron effective action. 
Such effects stem from the real-time production of charged pions, 
and arise akin to the Schwinger mechanism%
~\cite{Schwinger:1951nm}.
The production rate is suppressed by an exponential factor, 
$\propto \exp ( - \mathcal{N} m_\pi^2 /  |\vec{E} | )$, 
where 
$\mathcal{N}$ 
is a pure number. 
This rate has been computed using chiral perturbation theory%
~\cite{Tiburzi:2008ma}, 
but does not affect Euclidean space lattice simulations. 
Indeed, 
the Schwinger mechanism is absent in Euclidean space, 
because the essential difference between Euclidean electric, 
$\vec{\mathcal{E}}$, 
and magnetic fields, 
$\vec{B}$,
is only the direction used to measure correlation functions. 
Consequently the perturbative terms in the Minkowski-space effective action can be matched to those in the Euclidean-space
effective action using a trivial analytic continuation, 
$\vec{E} \to i \vec{\mathcal{E}}$. 
A thorough discussion is contained in%
~\cite{Tiburzi:2008ma}. 
}
%
%
Terms in the effective action are constrained by the underlying symmetries:
gauge invariance, Lorentz invariance,
$\mathcal{C}$, 
$\mathcal{P}$, 
and 
$\mathcal{T}$. 
The leading operator is the magnetic moment term,
$\overline{N} \, \sigma_{\mu \nu} F^{\mu \nu} N$,
 and the corresponding effective Hamiltonian is given by:
$H_n^{(1)} 
= 
- \mu_n  \, \vec{K} \cdot \vec{E}$,
where the matrices 
$\vec{K}$ 
are generators of boosts in the spin-$1/2$ representation of the Lorentz group. 
Operators at the next order contain the 
$s$-wave and $d$-wave couplings of the neutron to two photons, 
$\overline{N} \, N  \, F_{\mu \nu} F^{\mu \nu} $,
and
$i ( \overline{N} \, \gamma_\mu \partial_\nu N - \partial_\nu \overline N \, \gamma_\mu N) 
F^{ \{ \mu \rho} F_{\rho} {}^{\nu \}}$.
While the latter term would na\"ively be supressed by a power of the neutron mass, 
a field redefinition is required to arrive at a canonically normalized kinetic term,
and both operators end up being of the same order.
A linear combination of these two operators gives rise to the second-order effective Hamiltonian:
$H_n^{(2)}
= 
- \frac{1}{2} \alpha_E \vec{E}^2$.

Both effective interactions, 
$H_n^{(1)}$ 
and 
$H_n^{(2)}$,
give rise to electric dipole moments (EDMs). 
The magnetic moment interaction generates a motional EDM%
~\cite{Einstein:1906aa},
$\vec{d} \, {}^{(1)} = \mu_n \, \vec{\sigma} \times \vec{v}$, 
where 
$\vec{v}$ 
is a small neutron velocity. 
The interaction energy of the dipole and the electric field is merely
$- \vec{d} \, {}^{(1)} \cdot \vec{E} = - \vec{\mu} \cdot \vec{B}$, 
where 
$\vec{B} = \vec{v} \times \vec{E}$ 
is the magnetic field in the neutron's rest frame. 
The induced EDM, 
$\vec{d} \, {}^{(2)} = - \alpha_E \vec{E}$, 
is proportional to the strength of the applied field with a coefficient 
that is the electric polarizability.

The electric polarizability lowers the neutron's energy. 
For a neutron at rest, 
the energy shift is merely 
$\Delta E = - \frac{1}{2} \alpha_E \vec{E}^2$. 
Less obvious,
however, 
is the effect of the magnetic moment. 
The leading-order contribution clearly vanishes for a neutron at rest. 
One must treat the magnetic moment operator to second order to account for all
terms in the neutron energy at 
$\mathcal{O} (\vec{E}^2)$.
For a small velocity 
$\vec{v}$, 
the second order contribution has the form
\begin{equation}
\propto
\mu_n ( \vec{v} \times \vec{E} ) \frac{1}{0 - \frac{1}{2} M_n \vec{v} \, {}^2 } \,  \mu_n ( \vec{v} \times \vec{E} )
,\end{equation}
and survives the 
$\vec{v} \to 0$ 
limit. 
In this way, 
one avoids the neutron pole. 
Beyond this schematic discussion, 
one can make the result exact by calculating the neutron propagator to all orders in 
$H_n^{(1)}$ 
and 
$H_n^{(2)}$. 
The result is a neutron propagator with an energy shift of the form
\begin{equation} \label{eq:DE}
\Delta E = - \frac{1}{2} \left( \alpha_E - \frac{\mu_n^2}{M_n} \right) \vec{E}^2
.\end{equation}
As a consequence, studying the electric-field dependence of unpolarized neutron correlation functions is not enough to determine
the electric polarizability without knowledge of the magnetic moment.

In order to determine 
$\alpha_E$
from two-point functions, 
one must perform the analogue of a Born subtraction. 
For the case of a magnetic field, 
the magnetic moment can be isolated by considering  
spin-projected correlation functions. 
One way to access magnetic moments in an electric field is, 
by analogy, 
to use boost-projected correlation functions. 
We use this terminology to refer to the spinor structure;
the neutron remains at rest throughout. 
With projection matrices,
$\mathcal{P}_\pm = \frac{1}{2} ( 1  \pm K_3 )$, 
we observe that the boost-projected two-point functions have the form
\begin{equation} 
\label{eq:Boost}
\texttt{Tr}
\left[
\mathcal{P}_\pm 
G( t )
\right]
= 
Z ( 1 \pm \mu_n E )
e^{ - i t  ( M_n + \Delta E) }
,\end{equation}
for the electric field
$\vec{E} = E \hat{z}$,
with the energy shift 
$\Delta E$ 
given in Eq.~(\ref{eq:DE}). 
Thus a simultaneous measurement of both boost-projected correlation functions 
will allow one to determine the magnetic moment and electric polarizability.%
\footnote{
As the lattice correlation functions are determined in Euclidean space, 
it is useful to quote the Euclidean version of Eq.~(\ref{eq:Boost}):
$\texttt{Tr}
\left[
\mathcal{P}_\pm 
G( t )
\right]
= 
Z ( 1 \pm \mu_n \mathcal{E} )
\exp [ - t  ( M_n + \Delta E) ]$, 
where 
$t$ is now the Euclidean time,
$\mathcal{P}_\pm$ are the ``boost'' generators in Euclidean space, 
and 
$\Delta E = + \frac{1}{2} \left( \alpha_E - \frac{\mu_n^2}{M_n} \right) \mathcal{E}^2$. 
}

While we have focused our discussion on the neutron, 
a similar analysis is also possible for the proton. 
One might be worried that proton observables would be inaccessible 
in background electric fields because energy is not a good quantum number. 
This is not a fundamental setback, 
however, 
as one need not work with states of good energy. 
We suggested the approach to handle charged particles in electric fields
using single-particle effective actions%
~\cite{Detmold:2006vu}. 
One must sum the Born-level couplings to the total charge
to arrive at the expected form of the charged particle correlation function. 
For the proton in an electric field, 
there is an additional Born coupling to the anomalous magnetic moment, 
and finally a non-Born term 
(which is the electric polarizability). 
Nonetheless, 
the utilization of boost-projected proton correlation functions 
allows one to access the anomalous magnetic moment and electric polarizability. 
The correlation functions do not have simple exponential behavior in time, 
but can be determined in a model-independent fashion.

\section{Nucleon in Electric Fields on a Lattice}

We now focus on our Euclidean space lattice calculations in background electric fields. 
To implement a background electric field, 
we can choose either a compact or non-compact formulation. 
We utilize the former in order to arrive at uniform fields%
~\cite{'tHooft:1979uj,Smit:1986fn}.
Our compact 
$U(1)$
gauge field is implemented with links%
\footnote{
Currently our simulations are restricted to post-multiplied background fields. 
As a consequence, 
physical predictions can only be made for the isovector nucleon magnetic moment, 
although the isovector polarizabilities are expected to be fairly insensitive to the 
sea quark charges.  
}
\begin{equation}
U_\mu(x)
= 
\exp( - i q \mathcal{E} x_4 \delta_{\mu, 3})
\exp( i q  \mathcal{E} T x_3 \delta_{\mu, 4} \delta_{x_4, T-1} )
.\end{equation}
Notice the links are unitary as mandated by 
$\mathcal{C}$
invariance, 
i.e.~the parameter 
$\mathcal{E}$ 
is real valued. 
With 
$q \mathcal{E} = 2 \pi n /  L T$, 
the electric field through each elementary plaquette of the lattice is uniformly 
$\mathcal{E}$ in the 
$x_3$-direction.
Some time ago, 
we investigated the effects of electric field gradients by using non-quantized fields%
~\cite{Detmold:2008xk}.
While energy shifts to neutral particle correlators were found to be on the percent level, 
this is undesirable because the expected energy shifts due to polarizabilities are the same size.

\begin{table}[t]
\caption{\label{t:lattice} 
Lattices, 
propagator inversions, 
and background field strengths used.  
Further details of the lattice action can be found in 
Refs.~\cite{Edwards:2008ja,Lin:2008pr} 
with 
$N_s \rightarrow 20$ 
for this work.  
The electric field strength is listed in terms of the integer $n$ appearing in the quantization condition. 
}
\begin{center}
\begin{tabular}{cccc|cc}
$\quad  N_s \quad $& $ \quad N_t \quad $& $\quad a_t m_l \quad $& $\quad a_t m_s \quad $& $\quad  m_\pi \quad $& $\quad  m_K \quad $	\\
\hline \hline
$20$& $128$ & $-0.0840$ & $-0.0743$& $ 390\, \texttt{MeV}$ & $546 \, \texttt{MeV}$ \\
\hline\hline
\tabularnewline
\end{tabular}
\begin{tabular}{c|ccccc}
&  $\quad n=0 \quad $& $\quad  n=\pm1 \quad $& $\quad  n=\pm2 \quad $& $\quad n=\pm3 \quad $& $\quad  n=\pm4 \quad $ \\
\hline \hline
$|e \, a_t a_s \mathcal{E}|  $& $0.00000$ & $0.00736$ & $0.01472$ & $0.02209$ & $0.02945$   \\
$N_{\texttt{src}}\times N_{\texttt{cfg}}$& $20\times200$& $20\times200$& $10\times200$& $10\times200$& $10\times200$\\
\hline \hline
\end{tabular}
\end{center}
\end{table}

To demonstrate our method of computing nucleon magnetic moments and electric polarizabilities, 
we used an ensemble of anisotropic clover lattices generated by the Hadron Spectrum Collaboration. 
Table~\ref{t:lattice} summarizes the lattices used in our computation. 
Because these lattices were designed for the investigation of excited-state hadrons, 
we utilize two-state fits in order to stabilize the extraction of ground-state parameters. 
We provide a survey of our results%
~\cite{Detmold:2010dx} 
focusing on three cases:
unpolarized neutron, boost-projected neutron, and boost-projected proton.

\begin{figure}
\begin{center}
\includegraphics[width=13cm]{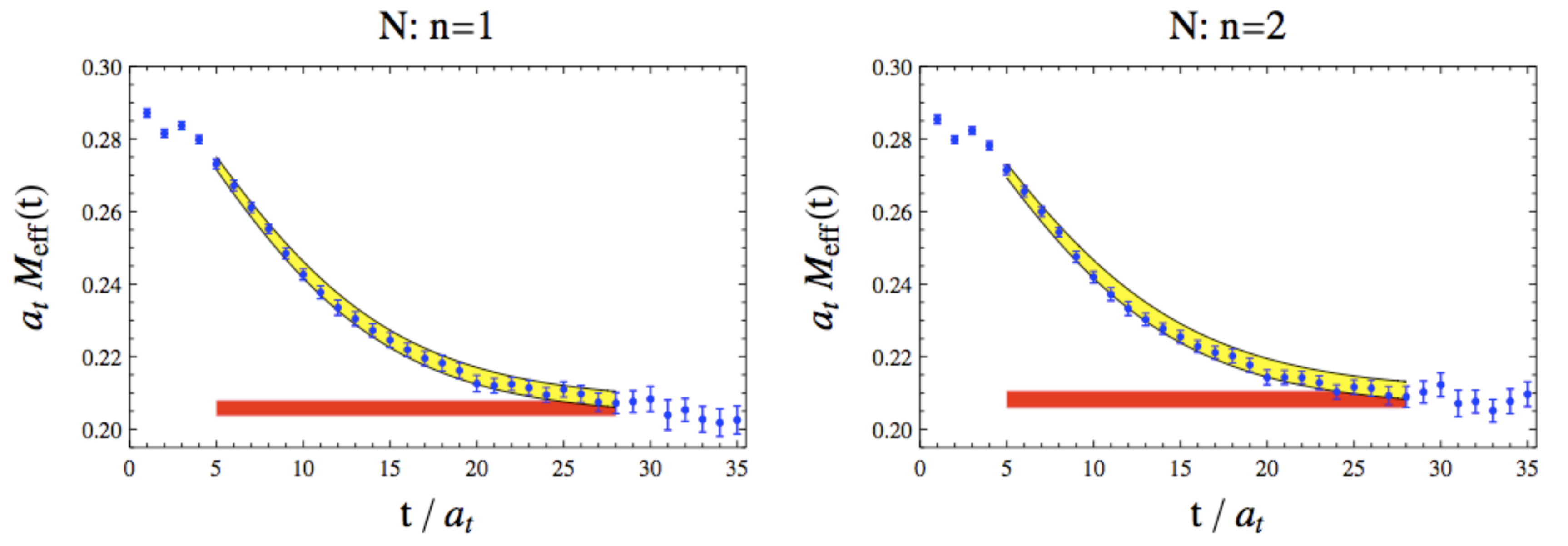}
\end{center}
\caption{\label{f:Nunpol} 
Effective mass plots for unpolarized neutron correlators, 
$\texttt{Tr} [ G_n(t) ]$. 
For two values of the electric field 
(corresponding to 
$n=1$ 
and 
$n=2$), 
we show the effective mass of the neutron correlator along with 
the effective mass of our two-state fit to the data with uncertainty band. 
The extracted ground-state energy is shown as a (red) flat band.  
}
\end{figure}

For unpolarized neutrons, 
we can only determine a combination of magnetic moment and electric polarizability, Eq.~(\ref{eq:DE}). 
To do this, 
we measure the unpolarized neutron correlation function for several values of the external electric field. 
Typical such measurements are shown in Fig.~\ref{f:Nunpol}. 
%
%
By measuring the neutron energy as a function of the applied field 
$\mathcal{E}$, 
we can then extract the coefficient of the 
$\mathcal{E}^2$
term. 
This is not simply the polarizability, 
rather the combination 
$\mathcal{A}_n = \frac{1}{2} \left( \alpha_E -  \frac{\mu_n^2}{ M_n} \right)$. 
Using the unpolarized neutron data, 
we find
$\mathcal{A}_n = 1.3 (9)(1)(1) \times 10^{-4} \, \texttt{fm}^3$.

On the other hand, 
with the boost-projected correlation functions, 
we can separate out the neutron magnetic moment
and electric polarizability. 
This is accomplished by measuring 
$\texttt{Tr} [ \mathcal{P}_\pm G_n(t) ]$
correlators for various values of the applied field
(the strength of which is indexed by 
$n$), 
and then performing simultaneous fits for each $n$. 
\begin{figure}
\begin{center}
\includegraphics[width=13cm]{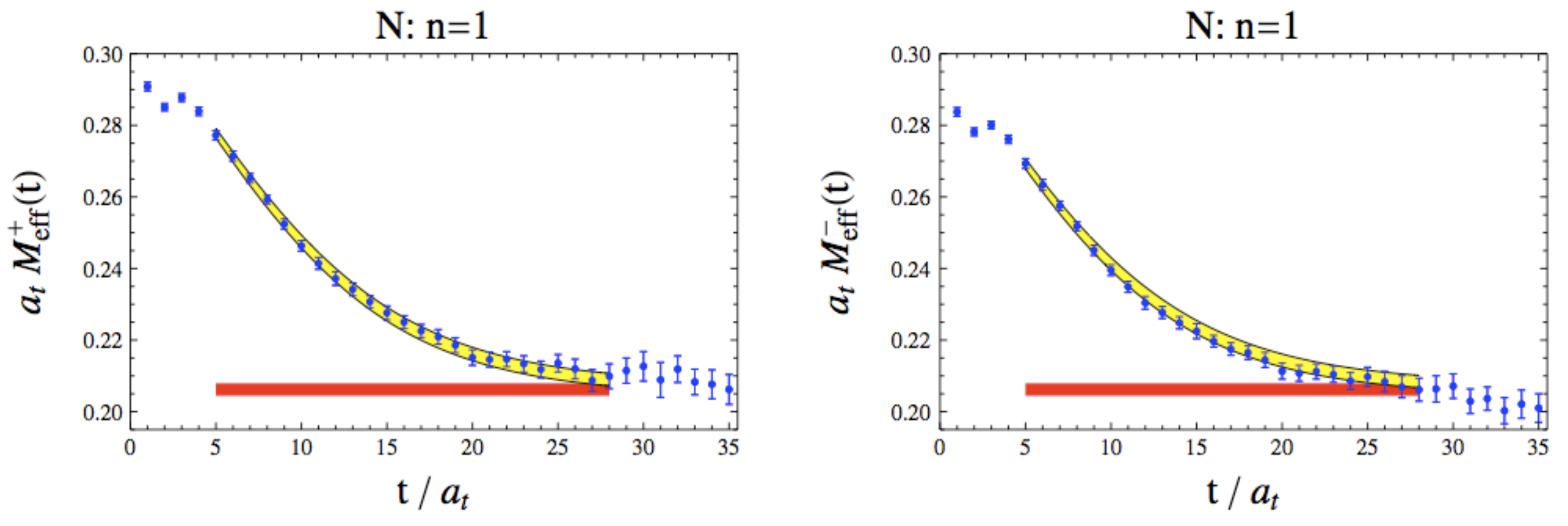}
\end{center}
\caption{\label{f:Nboost} 
Effective mass plots for boost-projected neutron correlators, 
$\texttt{Tr} [ \mathcal{P}_\pm G_n(t) ]$.  
For a single value of the electric field 
(corresponding to 
$n=1$), 
we show the two effective masses of boost-projected neutron correlators.
For these two plots,
we perform a simultaneous two-state fit in order to extract the neutron magnetic moment and electric polarizability.  
Results of the fit are also shown on the effective mass plots. 
The extracted ground-state energy is also shown as a (red) flat band.  
}
\end{figure}
%
%
Such fits allow one to determine the energy shift,
$\Delta E = \mathcal{A}_n \mathcal{E}^2 + \ldots$,
and the amplitudes 
$Z ( 1 \pm \mu_n \mathcal{E} )$. 
From this information, 
one can determine the coefficient 
$\mathcal{A}_n$
of the 
$\mathcal{E}^2$ 
term in the energy shift. 
We find
$\mathcal{A}_n = 1.3(7)(2)(1) \times 10^{-4} \, \texttt{fm}^3$, 
which is consistent with our analysis of the unpolarized correlators. 
Unlike that case, 
however, 
we can go further and determine the magnetic moment, 
$\mu_n = - 1.63 (10)(4)(5) \, [ \mu_N]$, 
and electric polarizability, 
$\alpha^n_E = 3.3(1.5)(2)(3) \times 10^{-4} \, \texttt{fm}^3$. 
It should be emphasized that these are ``connected'' values.

Finally we can perform a similar analysis for the boost-projected proton correlation functions. 
The time-dependence of the proton correlation functions is considerably different than that 
of the neutron because of the Born couplings to the total charge.  
\begin{figure}
\begin{center}
\includegraphics[width=13cm]{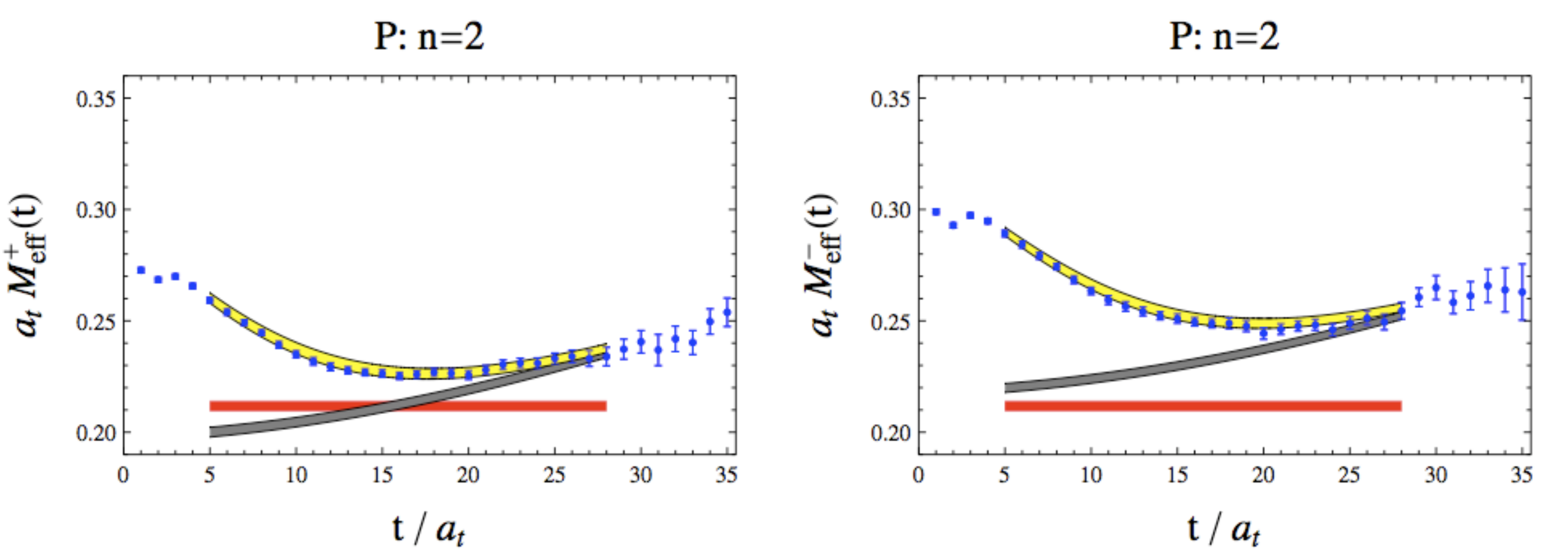}
\end{center}
\caption{\label{f:Pboost} 
Effective mass plots for boost-projected proton correlators,
$\texttt{Tr} [ \mathcal{P}_\pm G_n(t) ]$. 
For a single value of the electric field
(corresponding to 
$n=2$),
we plot the two effective masses of the boost-projected correlators. 
The proton correlator is not expected to have a simple exponential behavior at long times. 
The two effective masses of our simultaneous fit to both boost-projected correlators are shown, 
as well as just the ground-state contribution (gray) and ground-state rest energy (red). 
}
\end{figure}
%
%
While there is no simple exponential falloff at long times, 
there is a model-independent prediction for the behavior of the correlator.
Using the form predicted for a spin-1/2 particle, 
we can extract the magnetic moment and electric polarizability by performing
simultaneous fits to both boost-projected proton correlators. 
In Fig.~\ref{f:Pboost}, 
we display the behavior of the boost-projected proton correlators calculated in a particular electric field strength. 
From the fits over the various field strengths, 
we determine
$\mu_p = 2.63 (13)(1)(4) \, [\mu_N]$, 
and 
$\alpha^p_E = 2.4(1.9)(3)(2) \times 10^{-4} \, \texttt{fm}$. 
These are again only ``connected'' contributions.

\section{Outlook}

We have outlined our progress in the computation of hadronic electromagnetic properties using the background field method. 
In background electric fields, 
we showed that nucleon correlation functions depend on the magnetic moment and electric polarizability. 
Unpolarized neutron correlators do not allow one access to the electric polarizability, 
only the combination 
$\Delta E$
in Eq.~(\ref{eq:DE}). 
By contrast, 
boost-projected correlation functions allow separate determination of magnetic moments and electric polarizabilities. 
Using an ensemble of anisotropic gauge configurations 
(courtesy of the Hadron Spectrum Collaboration), 
we demonstrated our method successfully for both the neutron and proton. 
The proton had yet to be treated in background electric fields.

There are many refinement possible to our computation. 
The usual caveats must be issued: 
we have computed quantities at only one value of the pion mass, 
and one lattice spacing. 
Due to computational restrictions, 
the background field was not included in the gauge field generation. 
Additionally there are non-standard finite volume effects that one must worry about%
~\cite{Hu:2007eb,Tiburzi:2008pa,Detmold:2009fr}.
For example, 
the uniform electric field does not, strictly speaking, lead to a periodic lattice action. 
The action is only periodic up to a gauge transformation. 
Charged particle correlation functions experience a gauge defect at the edge of the lattice. 
This, 
however, 
is a calculable effect.
One can determine the effect of the defect on the two-point function by using a compact single-particle effective action. 
Another finite volume oddity concerns the non-trivial holonomy of the background field. 
New interactions are generated from virtual particles wrapping around the lattice. 
These should be exponentially small, moreover their effects can be addressed in an effective field theory framework. 
Work is currently underway to address these issues.

\begin{acknowledgments}
Work supported in part by 
Jefferson Science Associates, LLC under 
U.S.~Dept.~of Energy contract No.~DE-AC05-06OR-23177 (W.D.), 
and
the 
U.S.~Dept. of Energy, 
under
Grant Nos.~DE-SC000-1784 (W.D.),
~DE-FG02-93ER-40762 (B.C.T.), and
~DE-FG02-07ER-41527 (A.W.-L.).
\end{acknowledgments}



\end{document}